\input{aipcheck}

\documentclass[
    ,final            % use final for the camera ready runs
  ]
  {aipproc}

\layoutstyle{6x9}

\def\mueff{\mu_\mathrm{eff}}

\def\lamt{\Lambda_t}
\def\cta{\cos\theta_A}
\def\sta{\sin\theta_A}
\def\mupsilon{M_\Upsilon}
\def\tauptaum{\tau^+\tau^-}
\def\brups{\br(\Upsilon\to \gam\ai)}
\def\mupsilon{M_\Upsilon}
\def\mups{\mupsilon}

\def\mhusq{m_{H_u}^2}
\def\mhdsq{m_{H_d}^2}

\def\mqsq{m_Q^2}
\def\musq{m_U^2}
\def\mdsq{m_D^2}

\def\ls#1{\ifmath{_{\lower1.5pt\hbox{$\scriptstyle #1$}}}}
\def\lss#1{\ifmath{^{\,\lower2.5pt\hbox{$\scriptstyle #1$}}}}

\def\call{{\cal L}}

\def\mstopbar{\overline m_{\wtil t}}

\def\what{\widehat}

\def\del{\delta}

\def\mx{M_X}

\def\mz{m_Z}

\def\mhi{m_{\hi}}

\def\h{h}
\def\mh{m_{\h}}

\def\lam{\lambda}
\def\kap{\kappa}
\def\alam{A_\lam}
\def\akap{A_\kap}

\def\call{{\cal L}}

\def\wtil{\widetilde}
\def\what{\widehat}
\def\tauptaum{\tau^+\tau^-}

\def\lsim{\mathrel{\raise.3ex\hbox{$<$\kern-.75em\lower1ex\hbox{$\sim$}}}}
\def\gsim{\mathrel{\raise.3ex\hbox{$>$\kern-.75em\lower1ex\hbox{$\sim$}}}}
\def\ifmath#1{\relax\ifmmode #1\else $#1$\fi}
\def\half{\ifmath{{\textstyle{1 \over 2}}}}

\def\third{\ifmath{{\textstyle{1 \over 3}}}}

\def\vev#1{\langle #1 \rangle}
\def\lam{\lambda}

\def\mplanck{M_{\rm P}}
\def\mpl{\mplanck}

\def\mhi{m_{h_1^0}}

\def\emiss{/\hskip-9pt E}

\def\calo{{\cal O}}
\def\eg{{\it e.g.}}

\def\stopone{\wt t_1}
\def\stoptwo{\wt t_2}

\def\mstopone{m_{\stopone}}
\def\mstoptwo{m_{\stoptwo}}
\def\mstopmean{\sqrt{\mstopone\mstoptwo}}

\def\calo{{\cal O}}
\def\eg{{\it e.g.}}

\def\mstopone{m_{\stopone}}

\def\hsm{h_{\rm SM}}
\def\mhsm{m_{\hsm}}
\def\hl{h^0}
\def\hh{H^0}

\def\mhl{m_{\hl}}
\def\mhh{m_{\hh}}

\def\tanb{\tan\beta}

\def\mt{m_t}
\def\mb{m_b}
\def\mtau{m_\tau}
\def\mz{m_Z}
\def\mw{m_W}
\def\mgut{M_U}
\def\mx{M_X}

\def\cnone{\wt\chi^0_1}

\def\cntwo{\wt\chi^0_2}

\def\wt{\widetilde}

\def\MPL #1 #2 #3 {{\sl Mod.~Phys.~Lett.}~{\bf#1} (#3) #2}
\def\NPB #1 #2 #3 {{\sl Nucl.~Phys.}~{\bf #1} (#3) #2}
\def\PLB #1 #2 #3 {{\sl Phys.~Lett.}~{\bf #1} (#3) #2}
\def\PR #1 #2 #3 {{\sl Phys.~Rep.}~{\bf#1} (#3) #2}
\def\PRD #1 #2 #3 {{\sl Phys.~Rev.}~{\bf #1} (#3) #2}
\def\PRL #1 #2 #3 {{\sl Phys.~Rev.~Lett.}~{\bf#1} (#3) #2}
\def\RMP #1 #2 #3 {{\sl Rev.~Mod.~Phys.}~{\bf#1} (#3) #2}
\def\ZPC #1 #2 #3 {{\sl Z.~Phys.}~{\bf #1} (#3) #2}
\def\IJMP #1 #2 #3 {{\sl Int.~J.~Mod.~Phys.}~{\bf#1} (#3) #2}
\def\NIM #1 #2 #3 {{\sl Nucl.~Inst.~and~Meth.}~{\bf#1} {#3} #2}
\def\lam{\lambda}
\def\br{B}
\def\tauptaum{\tau^+\tau^-}

\def\gam{\gamma}

\def\anti{\overline}
\def\epem{e^+e^-}

\def\ie{{\it i.e.}}
\def\eg{{\it e.g.}}

\def\anti{\overline}

\def\ai{a_1}

\def\mai{m_{\ai}}

\def\fbi{~{\rm fb}^{-1}}

\def\mev{~{\rm MeV}}
\def\gev{~{\rm GeV}}
\def\tev{~{\rm TeV}}
\def\mt{m_t}
\def\mb{m_b}

\def\hi{\h_1}
\def\hii{\h_2}

\def\mhi{m_{\hi}}

\newcommand{\nc}{\newcommand}
\nc{\beq}{\begin{equation}}   \nc{\eeq}{\end{equation}}
\nc{\bea}{\begin{eqnarray}}   \nc{\eea}{\end{eqnarray}}
\nc{\baa}{\begin{array}}      \nc{\eaa}{\end{array}}
\nc{\bit}{\begin{itemize}}    \nc{\eit}{\end{itemize}}
\nc{\ben}{\begin{enumerate}}  \nc{\een}{\end{enumerate}}
\nc{\bce}{\begin{center}}     \nc{\ece}{\end{center}}
\def\beqa{\begin{eqnarray}}
\def\eeqa{\end{eqnarray}}
\def\bed{\begin{description}}
\def\eed{\end{description}}
\def\bmini{\begin{minipage}}
\def\emini{\end{minipage}}
\newcommand{\ba}{\begin{array}}
\newcommand{\ea}{\end{array}}

\def\calo{{\cal O}}
\def\eg{{\it e.g.}}

\def\tanb{\tan\beta}

\def\ie{{\it i.e.}}
\def\what{\widehat}

\def\call{{\cal L}}
\def\half{{1\over 2}}
\def\del{\delta}
\def\gam{\gamma}

\def\lam{\lambda}

\def\calo{{\cal O}}

\def\anti{\overline}

\def\vev#1{{\langle #1 \rangle}}

\def \fbi{~{\rm fb^{-1}}} 
\def\simle{%  ``less than about'' symbol
    \mathrel{\rlap{\raise 0.511ex 
        \hbox{$<$}}{\lower 0.511ex \hbox{$\sim$}}}}

\def\slashchar#1{\setbox0=\hbox{$#1$}           % set a box for #1
   \dimen0=\wd0                                 % and get its size
   \setbox1=\hbox{/} \dimen1=\wd1               % get size of /
   \ifdim\dimen0>\dimen1                        % #1 is bigger
      \rlap{\hbox to \dimen0{\hfil/\hfil}}      % so center / in box
      #1                                        % and print #1
   \else                                        % / is bigger
      \rlap{\hbox to \dimen1{\hfil$#1$\hfil}}   % so center #1
      /                                         % and print /
   \fi}

\def\ups{\Upsilon}

\begin{document}

\title 
{New (and Old) Perspectives on Higgs Physics\footnote{This writeup is based on a presentation at Scadron 70,
    ``Workshop on Scalar Mesons and Related Topics'', Lisbon, Portugal,
    February, 2008.}}

\classification{12.60.Fr,12.60.Jv,12.60.-i,14.80.Cp,11.30.Pb}
\keywords{Document processing, Class file writing, \LaTeXe{}}

\author{John F. Gunion}{
  address={Department of Physics, University of California, Davis, CA 95616},
  email={jfgunion@ucdavis.edu},
  thanks={This writeup is based on a presentation at Scadron 70,
    Workshop on Scalar Mesons and Related Topics, Lisbon, Portugal,
    February, 2008.}
}

\begin{abstract}
  Old and new ideas regarding Higgs physics are reviewed. We first
  summarize the quadratic divergence / hierarchy problem which strongly
  suggests that the SM Higgs sector will be supplemented by new
  physics at high scales. We next consider means for delaying the
  hierarchy problem of the SM Higgs sector to unexpectedly high
  scales.  We then outline the properties of the most ideal Higgs
  boson. The main advantages of a supersymmetric solution to the high
  scale problems are summarized and the reasons for preferring the
  next-to-minimal supersymmetric model over the minimal supersymmetric
  model in order to achieve an ideal Higgs are emphasized.  This leads
  us to the strongly motivated scenario in which there is a Higgs $h$
  with SM-like $WW,ZZ$ couplings and $\mh\sim 100\gev$ that decays via
  $h\to aa$ with $m_a<2m_b$, where $m_a>2m_\tau$ is preferred,
  implying $a\to \tauptaum$.  The means for detecting an $h\to aa\to
  4\tau$ signal are then discussed. Some final cautionary and
  concluding remarks are given.

\end{abstract}

\date{\today}

\maketitle

\section{Introduction}

The number one issue in Higgs physics is the solution 
of the hierarchy / fine-tuning problems that arise in the Standard
Model and Higgs sector extensions thereof from quadratically divergent
one-loop corrections to the Higgs mass. In fact, this ``quadratic
divergence fine-tuning'' is only one of three fine-tunings that we
will discuss. The second kind of fine-tuning is that sometimes called
``electroweak fine-tuning''; it is the fine-tuning associated with
getting the value of $\mz$ correct starting from GUT-scale parameters
of some model that already embodies a solution to the quadratic
fine-tuning problem. A third type of fine-tuning will emerge in the context of
the next-to-minimal supersymmetric model solution to avoiding electroweak
fine-tuning.

\begin{figure}[h!]
\includegraphics[width=3in]{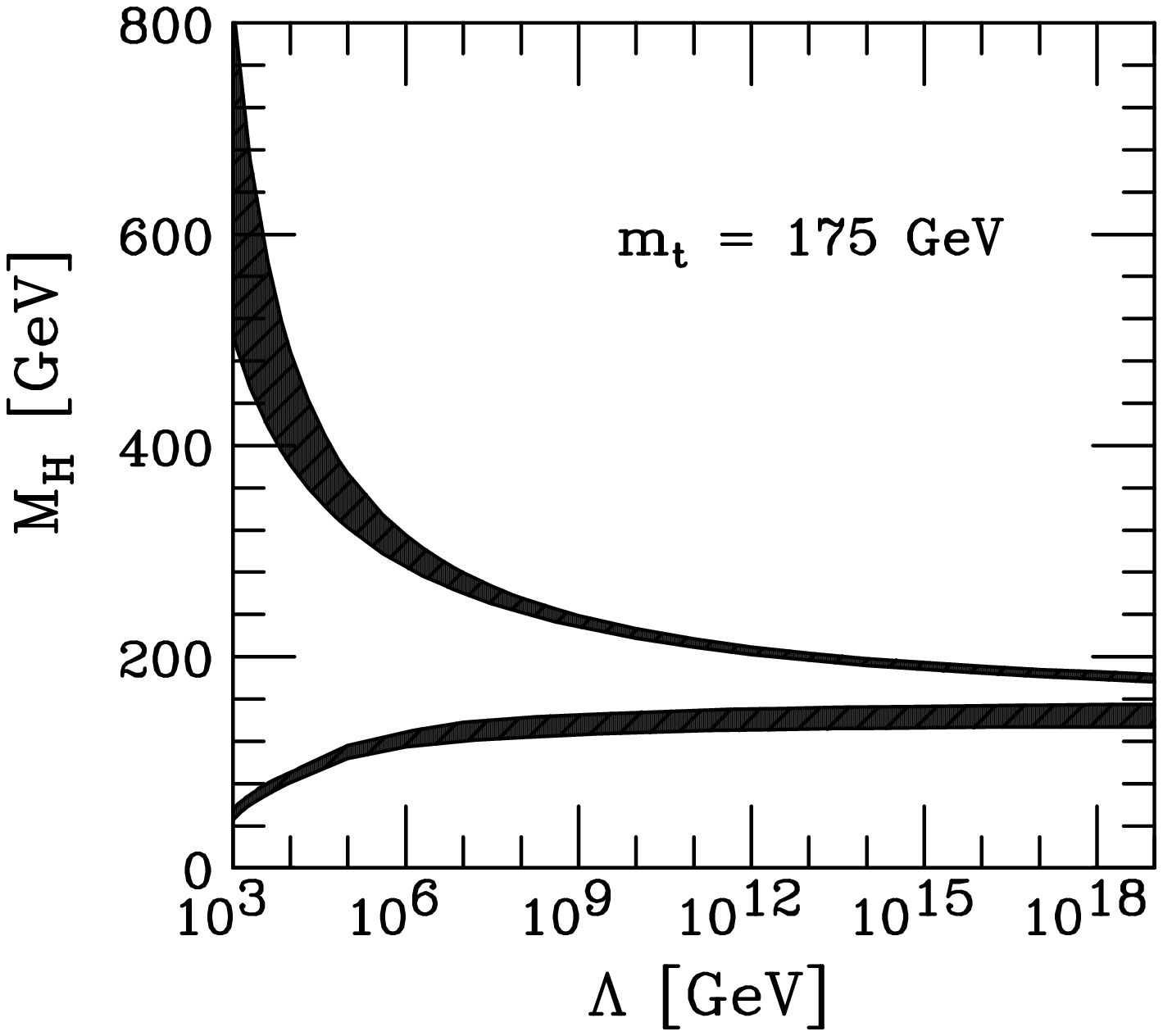}\includegraphics[width=3in]{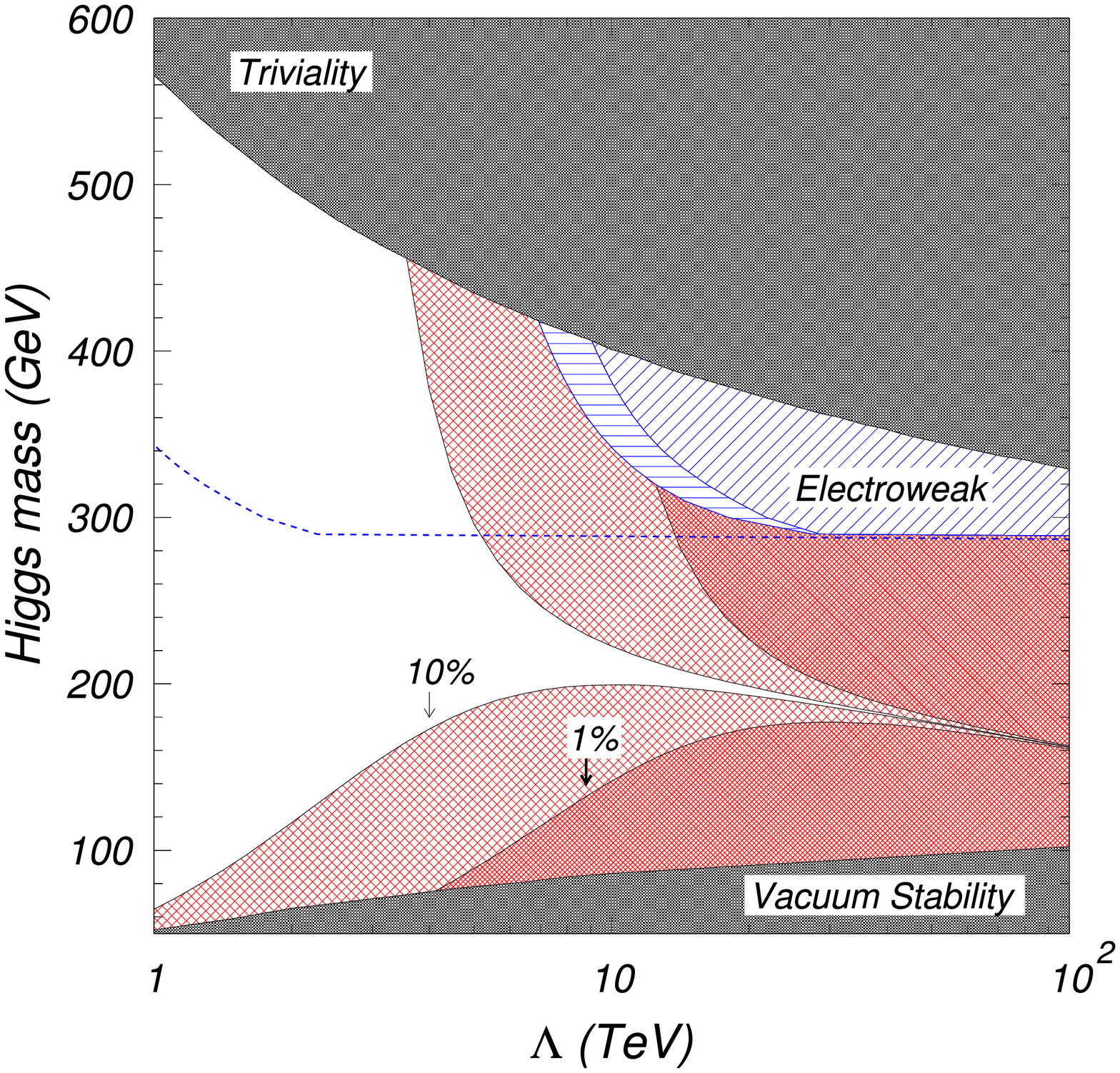}
\caption{Left: Triviality and global minimum constraints
on $\mhsm$ vs. $\Lambda$ from Ref.~\cite{Riesselmann:1997kg}. Right: Fine-tuning constraints on $\Lambda$, from Ref.~\cite{Kolda:2000wi}.}
\label{trivialityfinetuning}
\end{figure}

Were it not for the quadratic divergence fine-tuning problem, there is
nothing to forbid the SM from being valid all the way up to the Planck
scale.  The two basic theoretical constraints on $\mhsm$ as a function
of the scale $\Lambda$ at which new physics enters are:
\bit
\item the Higgs self coupling should not blow up below scale $\Lambda$
  --- this leads to an upper bound on $\mhsm$ as a function of $\Lambda$.
\item the Higgs potential should not develop a new minimum at large
  values of the scalar field of order $\Lambda$ --- this leads to a
  lower bound on $\mhsm$ as a function of $\Lambda$.  
\eit
The SM remains consistent with these two constraints all the way up to
$\Lambda\sim\mpl$ if $130\lsim\mhsm\lsim 180\gev$. This is shown in
Fig.~\ref{trivialityfinetuning}.
However, it is generally believed that the SM cannot be the full
theory all the way up to $\mpl$ due to quadratic divergence of loop
corrections to the Higgs mass. Because of this divergence, a light
Higgs is not ``natural'' in the SM context given the large
``hierarchy'' between the $100\gev$ and $\mplanck$ scales.  Assuming
that the SM is valid up to some large scale $\Lambda$, to obtain the
low Higgs mass favored by data (and required by $WW$ scattering
perturbativity) requires an enormous cancellation between top loop
corrections (as well as $W$, $Z$ and $\hsm$ loops) and the bare Higgs
mass of the Lagrangian. At one-loop, assuming
cutoff scale $\Lambda$,
\beq
 m_{\hsm}^2=m_0^2+{3\over 16\pi^2}
 (2\mw^2+\mz^2+\mhsm^2-4m_t^2)\Lambda^2 
\label{quaddiv}
\eeq 
where $m_0^2=2\lambda v_{SM}^2$ with $v_{SM}\sim 174\gev$. ($V\ni
\half \lam^2 [(\Phi^\dagger\Phi)^2 -v_{SM}^2 (\Phi^\dagger \Phi)]$ and
$\vev{\Phi}=v_{SM}$.)  Assuming no particular connection between the
contributions, we must fine tune $m_0^2$ to cancel the $\Lambda^2$
term with something like a precision of one part in $10^{32}$ if
$\Lambda=\mplanck$.  Further, this requires that the Higgs
self-coupling strength, $\lambda$, must be very large and
non-perturbative.  Keeping only the $m_t$ term with $\Lambda\to
\Lambda_t$, one measure of fine-tuning is:
\beq
F_t(\mhsm)=\left|{\partial\del\mhsm^2\over \partial \lamt ^2} {\lamt ^2\over
  \mhsm^2}\right|={3\over 4\pi^2}{m_t^2\over v_{SM}^2}{\lamt ^2\over \mhsm^2}\equiv K {\lamt^2\over\mhsm^2}\,.
\eeq
Given a maximum acceptable $F_t$, new physics must enter at or below
the scale
\begin{equation}
\lamt \lsim {2\pi v_{SM}\over \sqrt 3 m_t} \mhsm F_t^{1/2}\sim 400 \; \mbox{GeV} \left( {\frac{\mhsm}{115 \; \mbox{GeV}}}
\right){F_t^{1/2}}.
\label{eq:LSM}
\end{equation}
$F_t>10$, corresponding to fine-tuning parameters with a precision of
better than 10\%, is deemed problematical. For $\mhsm\sim 100\gev$, as
preferred by precision electroweak data, this implies new physics
somewhat below $1\tev$, in principle well within LHC reach.

\section{Options for Delaying New Physics}

Given that by definition new physics enters at scale $\Lambda$,
it is generically interesting to understand how the
quadratic divergence fine-tuning problem can be delayed to $\Lambda$
values substantially above $1\tev$, thereby making LHC new-physics
signals more difficult to detect. Two possible ways are the following. 
\ben
\item $\mhsm$ could obey the ``Veltman'' condition
  \cite{Veltman:1980mj} (see also \cite{Fang:1996cn} and
  \cite{Scadron:2006dy}),
\beq
\mhsm^2=4\mt^2-2\mw^2-\mz^2\sim (317\gev)^2\,,
\label{veltcond}
\eeq
for which the coefficient of $\Lambda^2$ in Eq.~(\ref{quaddiv})
vanishes. However, it turns out that at higher loop order, one must
carefully coordinate the value of $\mhsm$ with the value of $\Lambda$
\cite{Kolda:2000wi}.  Just as we do not want to have a fine-tuned
cancellation of the two terms in Eq.~(\ref{veltcond}), we also do not
want to insist on too fine-tuned a choice for $\mhsm$ (in the SM,
there is no symmetry that predicts any particular value).  The
right-hand plot of Fig.~\ref{trivialityfinetuning} shows the result
after taking this into account.  The upper bound for $\Lambda$ at
which new physics must enter is largest for $\mhsm\sim 200\gev$ where
the SM fine-tuning would be $10\%$ if $\Lambda\sim 30\tev$. At this
point, one would have to introduce some kind of new physics.  However,
we already know that there is a big problem with this approach --- the
latest $m_t$ and $m_W$ values when combined with LEP precision
electroweak data require $\mhsm<160\gev$ at 95\% CL.

\item An alternative approach to delaying quadratic divergence
  fine-tuning is to employ the multi-doublet model of
  \cite{Espinosa:1998xj}. In this model, the $ZZ$ coupling is shared
  among (perhaps many) Higgs mass eigenstates because the SM vev is
  shared among the corresponding Higgs fields.  A bit of care in
  setting the scenario up is needed to avoid seeing other Higgs while
  at the same time satisfying the precision EW constraint:
\bea \sum_i \frac{v_i^2}{v_{SM}^2} \ln m_{h_i} \lsim \ln
\left( 160 \gev \right), 
\eea 
where $\vev{\Phi_j}\equiv v_j$ and $\sum_j v_j^2=v_{SM}^2 \sim (175
\gev)^2$ .  If you don't want LEP to have seen any sign of a Higgs
boson, the PEW constraint can still be satisfied even if all the Higgs
decay in SM fashion, so long as the eigenstates are not too much below
$100\gev$ and not degenerate.  But, of course, with enough $h_j$
eigenstates, Higgs decays will not be SM-like given the proliferation
of $h_j\to h_i h_i$ and $h_j\to a_ia_i$ decays.  The combination of
such decays and weakened production rates for the individual Higgs
bosons would make Higgs detection very challenging at the LHC and
require a high-luminosity linear collider. 
Returning to the quadratic divergence issue, we note that in the simplest
  case where all $h_i$ fields have the same top-quark Yukawa, $\lam_t$
  in $\call=\lam_t \anti t h_i t$, each $h_i$ has its top-quark-loop
  mass correction scaled by $f_i^2\equiv {v_i^2\over v_{SM}^2}$ and
  one gets a significantly reduced $F_t$ for each $h_i$:
\beq
F_t^i=f_i^2 F_t(m_i)=K f_i^2{ \lamt^2\over m_i^2}.
\label{ftreduced}
\eeq
Thus, multiple mixed Higgs allow a much larger $\lamt$ for a given
maximum acceptable common $F_t^i$.  A model with $4$ doublets can have
$F_t^i<10$ for $\lamt$ up to $5\tev$.
\een 
One good feature of delaying new physics is that large $\lamt$ implies
that significant corrections to low-$E$ phenomenology from
$\lamt$-scale physics (\eg\ FCNC) are less likely.  However, in the
end, there is always going to be a $\Lambda$ or $\lamt$ for which
quadratic divergence fine-tuning becomes unacceptable.  Ultimately we
will need new physics.  So, why not have it right away (\ie\ at
$\Lambda \lsim 1\tev$) and avoid the above somewhat ad hoc games.
This is the approach of supersymmetry, which (unlike Little Higgs or
UED or ....) solves the hierarchy problem once and for all, \ie\ there
is no need for an unspecified ultraviolet completion of the theory. We
will return to supersymmetry momentarily.

\section{Criteria for an Ideal Higgs Theory}

Theory and experiment have led us to a set of
criteria for an ideal Higgs theory.  We list these below.

\bit
\item It should allow for a light Higgs boson without quadratic
  divergence fine-tuning.  
  
\begin{figure}
\includegraphics[width=4in,angle=0]{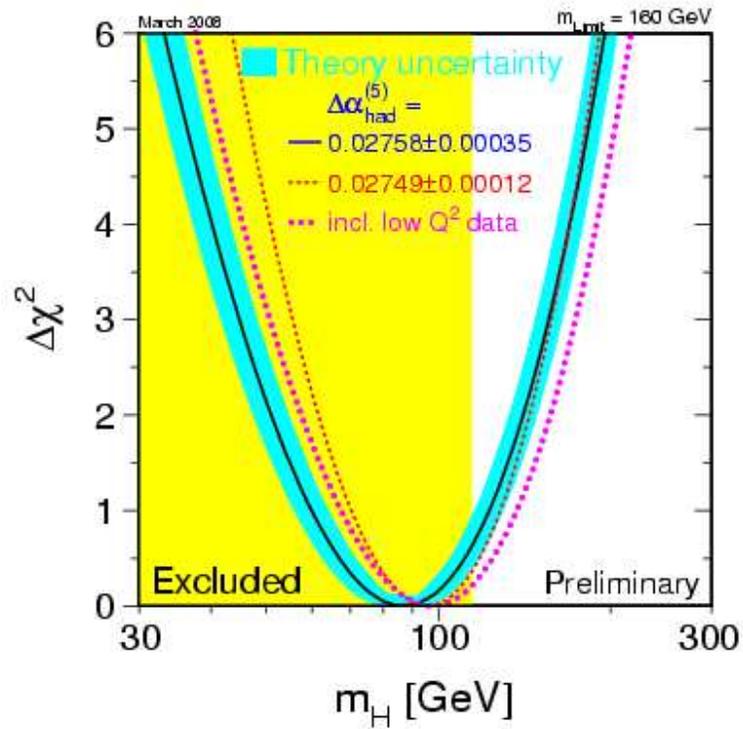}
\caption{The ``blue-band'' plot showing the preferred Higgs mass range
  as determined using precision electroweak data and measured top
  and $W$ boson masses.}
\label{blueband}
\end{figure}

\item It should predict a Higgs with SM couplings to $WW,ZZ$ and with
  mass in the range preferred by precision electroweak data. The LEPEWWG
  plot from winter 2008 is shown in Fig.~\ref{blueband}.
 At 95\% CL, $\mhsm<160\gev$
and the $\Delta\chi^2$ minimum is between $80\gev$ and $100\gev$.

\item Thus, in an ideal model, the Higgs should have mass no larger
  than $100\gev$.  But, at the same time, one must avoid the LEP
  limits on such a light Higgs. One generic possibility is for the
  Higgs decays to be non-SM-like. The limits on various Higgs decay
  modes from LEP are given in Table~\ref{lepmodes}, taken from
  Ref.~\cite{Chang:2008cw}.  From this table, we see that to have
  $m_H\leq 100\gev$ requires that the Higgs decays to one of the final
  three modes or something even more exotic.
\begin{table}
\small
\caption {LEP $m_H$ Limits for an $H$ with SM-like $ZZ$ coupling, but
  varying decays. \label{lepmodes} }
\bigskip
\footnotesize
\begin{tabular}{|c|c|c|c|c|c|c|c|}
\hline
Mode & SM modes & $2\tau$ or $2b$ {\it only} & $2j$ & $ WW^*+ZZ^*$ &  $\gam\gam$ & $\emiss$ &  $4e,4\mu,4\gam$  \cr
Limit (GeV) & $114.4$ & $115$ & $113$ & $100.7$ & $117$ & $114$ &  $114$?     \cr
\hline
\hline
Mode & $4b$ & $4\tau$ & any (e.g. $4j$) & $2f+\emiss$ & & & \cr
Limit (GeV) & $110$ & $86$ & $82$ & $90$?  & & & \cr
\hline
\end{tabular}
\end{table}

\begin{figure}
\includegraphics[height=3in,width=3in,angle=90]{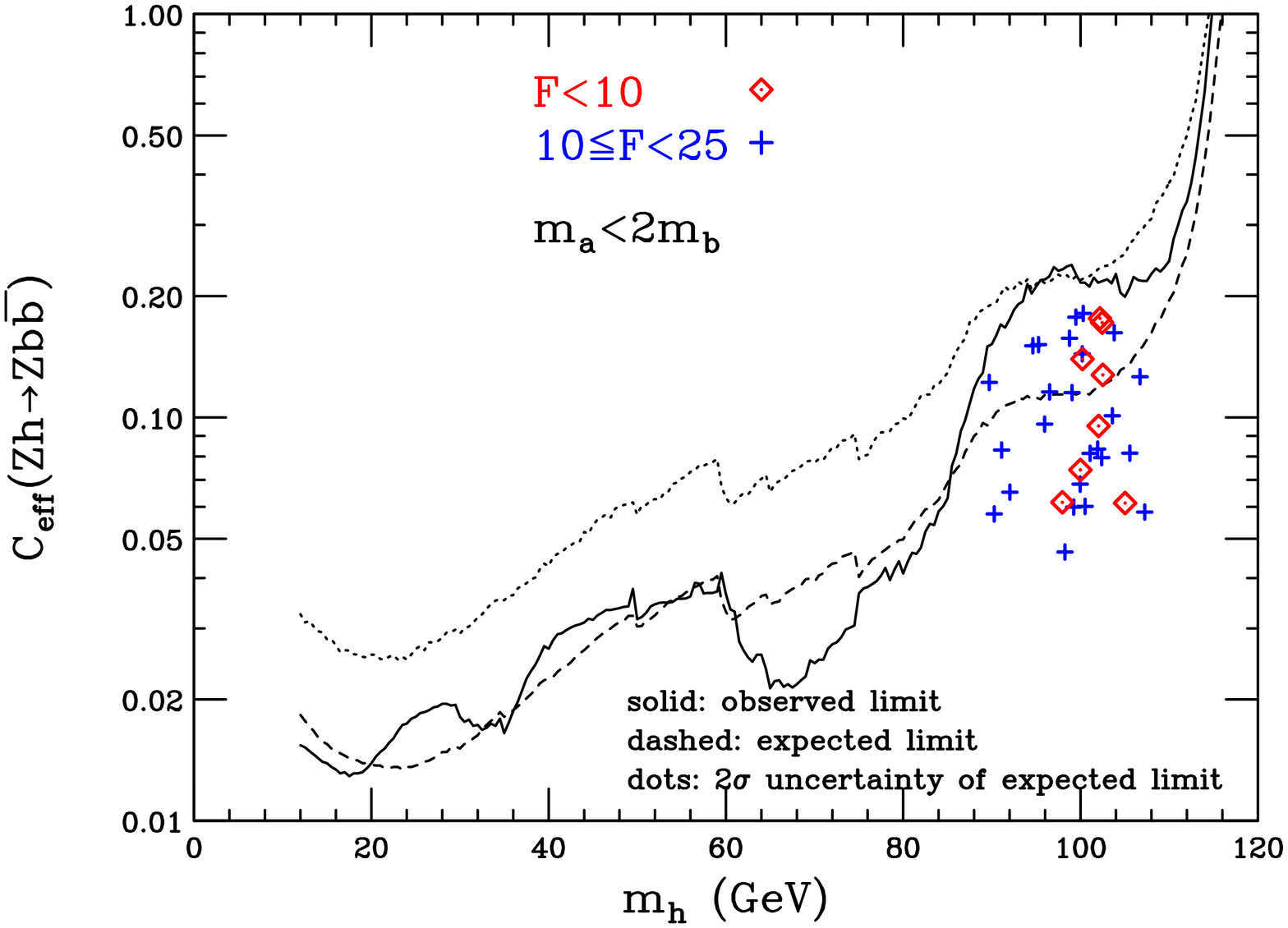}
\includegraphics[width=3in,angle=0]{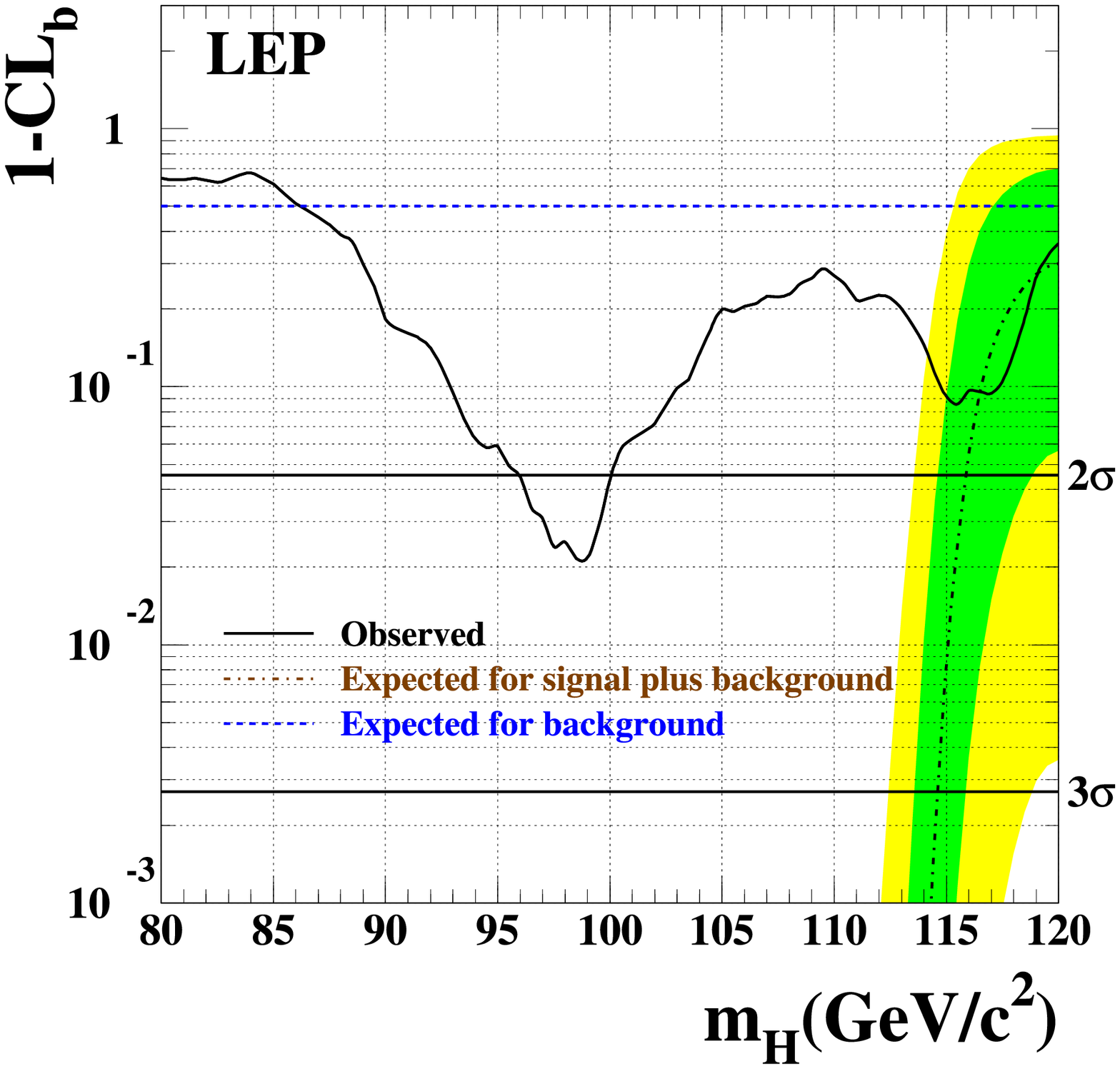}
\caption{LEP plots for the $Zb\anti b$ final state from the LEP Higgs
  Working Group.}
\label{clbplot}
\end{figure}

\item Perhaps the Higgs properties should be such as to predict the
  $2.3\sigma$ excess at $M_{b\anti b} \sim 98\gev$ seen in the
  $Z+b\anti b$ final state --- see Fig.~\ref{clbplot}.
For consistency with the observed excess, the $\epem\to Z H\to Z
b\anti b$ rate should be about one-tenth the SM value.  There are two
obvious ways to achieve this: (1) one could have $\br(H\to b\anti
b)\sim 0.1 \br(H\to b\anti b)_{SM}$ and $g_{ZZH}^2\sim g_{ZZ\hsm}^2$;
or (2) $\br(H\to b\anti b)$ could be SM-like but $g_{ZZH}^2\sim 0.1
g_{ZZ\hsm}^2$.

Regarding (1), almost any additional decay channel will severely
suppress the $b\anti b$ branching ratio.  A Higgs of mass 100 GeV has
a decay width into Standard Model particles that is only $2.6 \mev$,
or about $10^{-5}$ of its mass.  This implies that it doesn't take a
large Higgs coupling to some new particles for the decay width to
these new particles to dominate over the decay width to SM particles
--- see \cite{Gunion:1984yn}, \cite{Li:1985hy}, and
\cite{Gunion:1986nh} (as reviewed in \cite{Chang:2008cw}).  For
example, compare the decay width for $h\to b\anti b$ to that for $h\to
aa$, where $a$ is a light pseudoscalar Higgs boson.  Writing $\call\ni
g_{h aa}haa$ with $g_{haa}\equiv c\, {g m_h^2\over 2\mw}$ and ignoring
phase space suppression, we find
\bea 
{\Gamma(h\to
  aa)\over \Gamma(h\to b\anti b)}&\sim & 310\, c^2\left({m_{h}\over
    100\gev}\right)^2.
\eea
This expression includes QCD corrections to the $b\anti b$ width as
given in HDECAY which decrease the leading order $\Gamma(h\to b \anti
b)$ by about 50\%.  The decay widths are comparable for $c\sim 0.057$
when $m_h=100\gev$. Values of $c$ at this level or substantially
higher (even $c=1$ is possible) are generic in BSM models containing
an extended Higgs sector.

Regarding possibility (2), let us return to the scenario of
\cite{Espinosa:1998xj} in which the $ZZ$ coupling is shared among many
Higgs mass eigenstates.  To explain the $2.3 \sigma$ excess, there
should be a Higgs field having vev squared of order $0.1\times
v_{SM}^2$ and corresponding eigenstate with mass $\sim 100\gev$. (This
simple scenario assumes no Higgs mixing --- incorporation of mixing is
straightforward.)  An interesting special case is to construct a 2HDM
with $\mhl=98\gev$ and $g_{ZZ\hl}^2=0.1 g_{ZZ\hsm}^2$ and with
$\mhh=116\gev$ (the other LEP excess) and $g_{ZZ\hh}^2\sim 0.9
g_{ZZ\hsm}^2$ (see, for example, \cite{Drees:2005jg}).  As discussed
earlier, multiple Higgs games are also ``useful'' in that they can
delay the quadratic divergence fine-tuning problem to higher
$\Lambda$.

\eit

\section{Why Supersymmetry}

Ultimately, however, we must solve the quadratic divergence problem.
There are many reasons why supersymmetry is regarded as the leading
candidate for a theory beyond the SM that accomplishes this.  
Let us review them very briefly.
(a) SUSY is mathematically intriguing.
(b) SUSY is naturally incorporated in string theory.
(c) Elementary scalar fields have a natural place in SUSY, and so
  there are candidates for the spin-0 fields needed for electroweak
  symmetry breaking and Higgs bosons.
(d)
SUSY cures the naturalness / hierarchy problem (quadratic divergences
are largely canceled) in a particularly simple way.  And, it does so
without electroweak fine-tuning (see definition below) provided the
SUSY breaking scale is $\lsim 500\gev$. For example, the top quark
loop (which comes with a minus sign) is canceled by the loops of the
spin-0 partners called "stops" (which loops enter with a plus sign).
Thus, $\lamt^2$ is effectively replaced by $\mstopbar^2\equiv
\mstopone\mstoptwo$.

Overall, the most minimal version of SUSY, the MSSM comes close to
being very nice.  If we assume that all sparticles reside at the
$\calo(1\tev)$ scale and that $\mu$ is also $\calo(1\tev)$, then, the
MSSM has two particularly wonderful properties.
\begin{figure}[h!]
\includegraphics[width=2.5in]{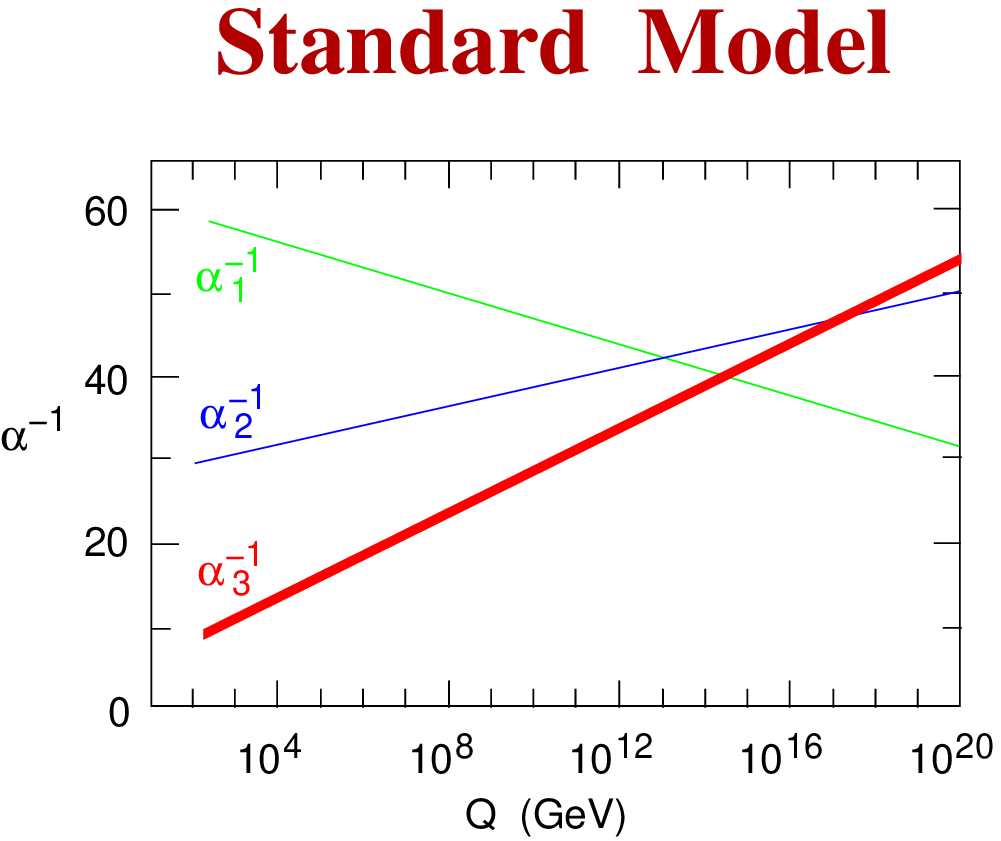}\includegraphics[width=2.5in]{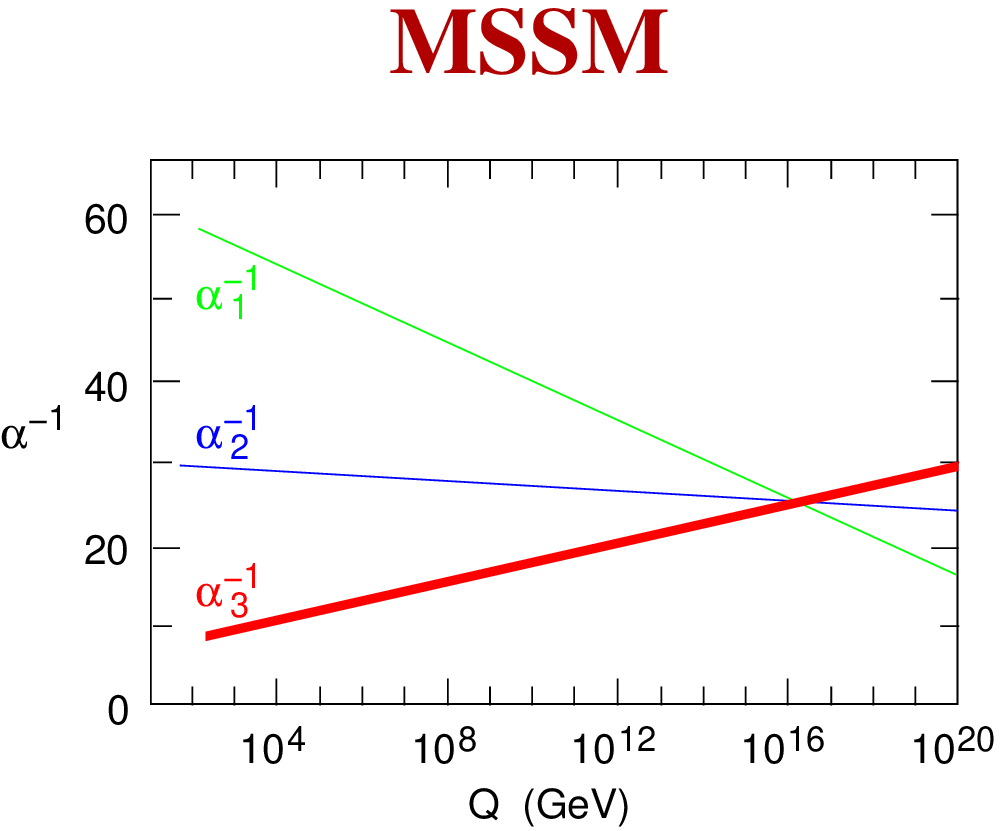}
\caption{Unification of couplings constants ($\alpha_i=g_i^2/(4\pi)$)
in the minimal supersymmetric model (MSSM) as compared to failure
without supersymmetry.}
\label{gaugeunification}
\end{figure}
First, the MSSM sparticle content plus two-doublet
Higgs sector leads to gauge coupling unification
at $\mgut\sim few\times 10^{16}\gev$, close to
 $\mplanck$ --- see Fig.~\ref{gaugeunification}. High-scale unification correlates well with the
attractive idea of gravity-mediated SUSY breaking.
\begin{figure}[h!]
\includegraphics[width=3in,height=3in]{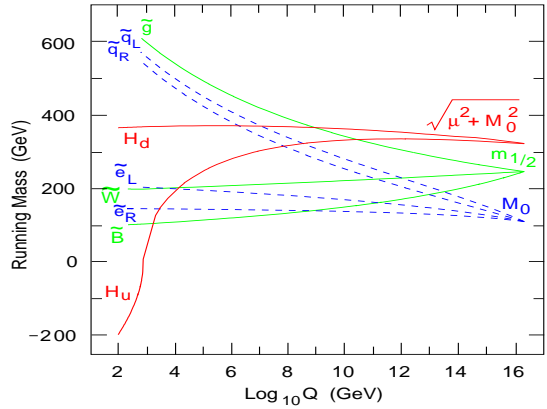}
\caption{Evolution of the (soft) SUSY-breaking masses or masses-squared,
  showing how $\mhusq$ is driven $<0$ at low $Q \sim \calo(\mz)$.}
\label{rewsb3}
\end{figure}
Second, starting with universal soft-SUSY-breaking masses-squared at
$\mgut$, the RGE's predict that the top quark Yukawa coupling will
drive one of the soft-SUSY-breaking Higgs masses-squared ($\mhusq$)
negative at a scale of order $Q\sim \mz$, thereby automatically
generating electroweak symmetry breaking
($\vev{H_u}=h_u,\vev{H_d}=h_d$, where $H_u$ and $H_d$ are the two
scalar Higgs fields of the MSSM) --- see Fig.~\ref{rewsb3}.  However,
as we shall discuss, fine-tuning of the GUT-scale parameters may be
required in order to obtain the correct value of $\mz$ unless, for
example, the stop masses are no larger than $2\mt$ or so.

\section{MSSM Problems}

However, the MSSM is suspect because of two critical problems.
\bit
\item {\bf The $\mu$ parameter problem:} In $W\ni \mu \what H_u \what
  H_d$,\footnote{Hatted (unhatted) capital letters denote superfields
    (scalar superfield components).} $\mu$ is dimensionful, unlike all
  other superpotential parameters.  Phenomenologically, it must be
  $\calo(1\tev)$ (as required for proper EWSB and in order that the
  chargino mass be heavier than the lower bounds from LEP and Tevatron
  experiments).  However, in the MSSM context the most natural values
  are either   $\calo(\mgut,\mplanck)$ or $0$.
  
\item {\bf LEP limits and Electroweak Fine-tuning:} Since the lightest
  Higgs, $h$, of the (CP conserving) MSSM has SM-like coupling {\it
    and} decays, the LEP limit of $\mh>114.4\gev$ applies for most of
  MSSM parameter space.  Such a $h$ is only possible for special MSSM
  parameter choices, for example large $\tanb=v_u/v_d$
  and large stop masses (roughly
  $\mstopmean\gsim 900\gev$) or large stop mixing. To quantify the
  problem we define
\beq
F={\rm Max}_p \left\vert {p\over \mz}{\partial\mz\over
      \partial p} \right\vert,
\eeq
where $p\in\left\{M_{1,2,3}, \mqsq, \musq, \mdsq, \mhusq, \mhdsq, \mu,
  A_t, B\mu,\ldots\right\}$ (all at $\mgut$).  These $p$'s are the
GUT-scale parameters that determine all the $\mz$-scale SUSY
parameters, and these in turn determine $v_{SM}^2$ to which $\mz^2$ is
proportional.  For example, $F>20$ means worse than $5\%$ fine-tuning
of the GUT-scale parameters is required to get the right value of
$\mz$, a level generally regarded as unacceptable.  Thus, an important
question is what is the smallest $F$ that can be achieved while
keeping $\mh>114\gev$.  The answer is (see, in particular,
\cite{Dermisek:2007ah,Dermisek:2007yt}): (a) For most of parameter
space, $F>100$ or so; (b) For a part of parameter space with large
mixing between the stops, $F$ can be reduced to $16$ at best ($6\%$
fine-tuning), but this part of parameter space has many other
peculiarities.  An ideal model would have $F\lsim 5$, which
corresponds to absence of any significant electroweak fine-tuning.
\eit

\section{The NMSSM}

Both problems are nicely solved by the next-to-minimal supersymmetric
model (NMSSM) in which a single extra singlet superfield is added to
the MSSM.  The new superpotential and associated soft-SUSY-breaking
terms are
\beq
W\ni \lam\what S\what H_u\what H_d+\third \kappa \what S^3\,,\quad
V\ni \lam\alam S H_u H_d+\third\kap\akap S^3\,.
\eeq
The explicit $\mu \what H_u \what H_d$ term found in the MSSM
superpotential is removed.  Instead, $\mu$ is automatically generated
by $\vev{S}\neq 0$ leading to $\mu_{eff}\what H_u \what H_d$ with
$\mu_{eff}=\lam \vev{S}$.  The only requirement is that $\vev{S}$ not
be too small or too large.  This is automatic if there are no
dimensionful couplings in the superpotential since $\vev{S}$ is then
of order the SUSY-breaking scale, which will be of order a $\tev$
or below.

\begin{figure}
\hspace*{-1.3cm}\includegraphics[width=4in,angle=90]{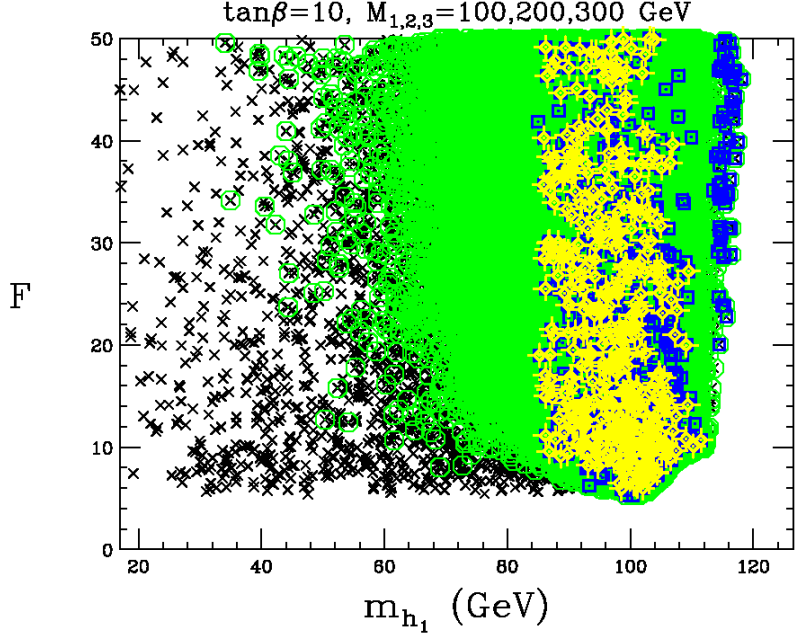}
\caption{$F$ vs. $\mhi$
  for $M_{1,2,3}=100,200,300\gev$ and $\tanb=10$. Small $\times$
  points have no constraints other than the requirement that they
  correspond to a global and local minimum, do not have a Landau pole
  before $\mgut$ and have a neutralino LSP. The O points are those
  which survive after stop and chargino mass limits are imposed, but
  no Higgs limits. The square points pass all LEP {\it single
    channel}, in particular $Z+2b$ and $Z+4b$, Higgs limits. The large
  yellow fancy crosses are those left after requiring $\mai<2\mb$, so
  that LEP limits on $Z+b's$, where $b's=2b+4b$, are not violated. }
\label{fvsmhinmssm}
\end{figure}

Electroweak fine-tuning and its implications for the NMSSM have been
studied in
\cite{Dermisek:2007ah,Dermisek:2007yt,Dermisek:2006py,Dermisek:2006ya,Dermisek:2006wr,Dermisek:2005gg,Dermisek:2005ar}
and reviewed in \cite{Accomando:2006ga,Chang:2008cw}.
Electroweak fine-tuning can be absent since the sparticles, especially
the stops, can be light without predicting a light Higgs boson with
properties such that it has already been ruled out by LEP, a point we
return to shortly.  A plot of $F$ as a function of the mass of the
lightest CP-even Higgs, $\mhi$, appears in Fig.~\ref{fvsmhinmssm}.
The electroweak fine-tuning parameter has a minimum of order $F\sim 5$
(which arises for stop masses of order 350 GeV) for $\mhi\sim
100\gev$, even without placing any experimental constraints on the
model (the $\times$ points).  This is perfect for precision
electroweak constraints because the $\hi$ has very SM-like $WW,ZZ$
couplings and an ideal mass.  However, most of the $\times$ points are
such that the $\hi$ is excluded by LEP.  Only the fancy-yellow-cross
points pass all LEP Higgs constraints, but there are many of these with
$F\sim 5$.  These points are such that $\mhi\sim 100\gev$ {\it and}
the $\hi$ avoids LEP Higgs limits by virtue of $\br(\hi\to
\ai\ai)>0.75$ with $\mai<2\mb$.  (Here, $\ai$ is the lightest of the
two CP-odd Higgs bosons of the NMSSM.)  In the $\hi\to \ai\ai\to
4\tau$ channel, the LEP lower limit is $\mhi>87\gev$.  In
the $\hi\to \ai\ai\to 4j$ channel, the LEP lower limit is
$\mhi>82\gev$ --- see Table~\ref{lepmodes}.

Further, there is an intriguing coincidence. For the many points with
$\br(\hi\to \ai\ai)>0.85$, then $\br(\hi\to b\anti b)\sim 0.1$ and the
$2.3\sigma$ LEP excess near $m_{b\anti b}\sim 98\gev$ in $\epem\to
Z+b's$ is perfectly explained.  There are a significant number of such
points in NMSSM parameter space.  For these points, the $\hi$
satisfies all the properties listed earlier for an ``ideal'' Higgs.
Further, for these points the GUT-scale SUSY-breaking parameters (such
as the Higgs soft masses-squared, the $\akap$ and $\alam$
soft-SUSY-breaking parameters, and the $A_t$ stop mixing parameter)
are particularly appealing being generically of the 'no-scale'
variety. That is, for the lowest $F$ points we are talking about,
almost all the soft-SUSY-breaking parameters are small at the GUT
scale.  This is a particularly attractive possibility in the string
theory context.

There is one remaining issue for these NMSSM scenarios.  We must ask
whether a light $\ai$ with the right properties is natural, or does
this require fine-tuning of the GUT-scale parameters?  This is the topic
of \cite{Dermisek:2006wr}.  The answer is that these scenarios can be
very natural.  First, we note that the NMSSM has a $U(1)_R$ symmetry
obtained when $\akap$ and $\alam$ are set to zero.  If this limit is
applied at scale $\mz$, then, $\mai=0$.  But, it turns out that then
$\br(\hi\to \ai\ai)\lsim 0.3$, which does not allow escape from the
LEP limit.  However, the much more natural idea is to impose the
$U(1)_R$ symmetry at the GUT scale.  Then, the renormalization group
often generates exactly the values for $\akap$ and $\alam$ needed to obtain
a light $\ai$ with large $\br(\hi\to\ai\ai)$.

\begin{figure}
\includegraphics[width=3in,angle=90]{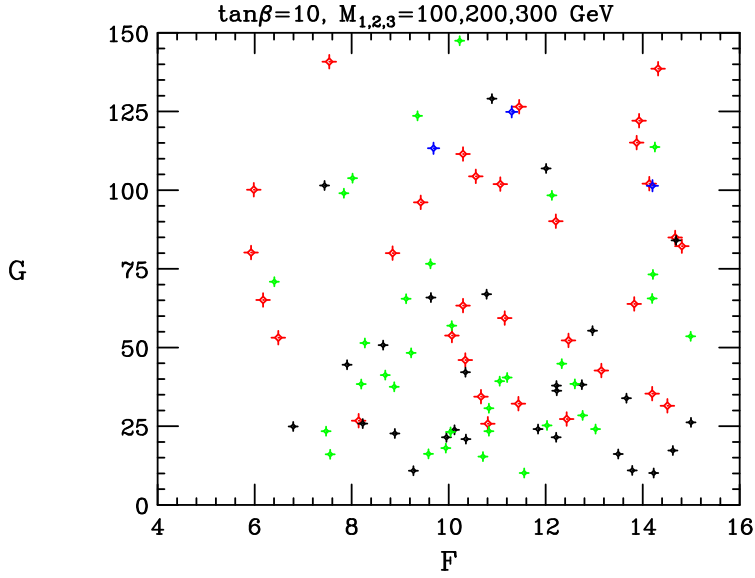}
\caption{$G$ vs. $F$
  for $M_{1,2,3}=100,200,300\gev$ and $\tanb=10$ for points with
  $F<15$ having $\mai<2\mb$ and large enough $\br(\hi\to\ai\ai)$ to
  escape LEP limits. The color coding is: blue = $\mai<2\mtau$; red
  $=2\mtau<\mai<7.5\gev$; green $=7.5\gev<\mai<8.8\gev$; and black $=
  8.8\gev<\mai<9.2\gev$.  }
\label{gvsf}
\end{figure}

\begin{figure}
\includegraphics[width=2.5in,angle=90]{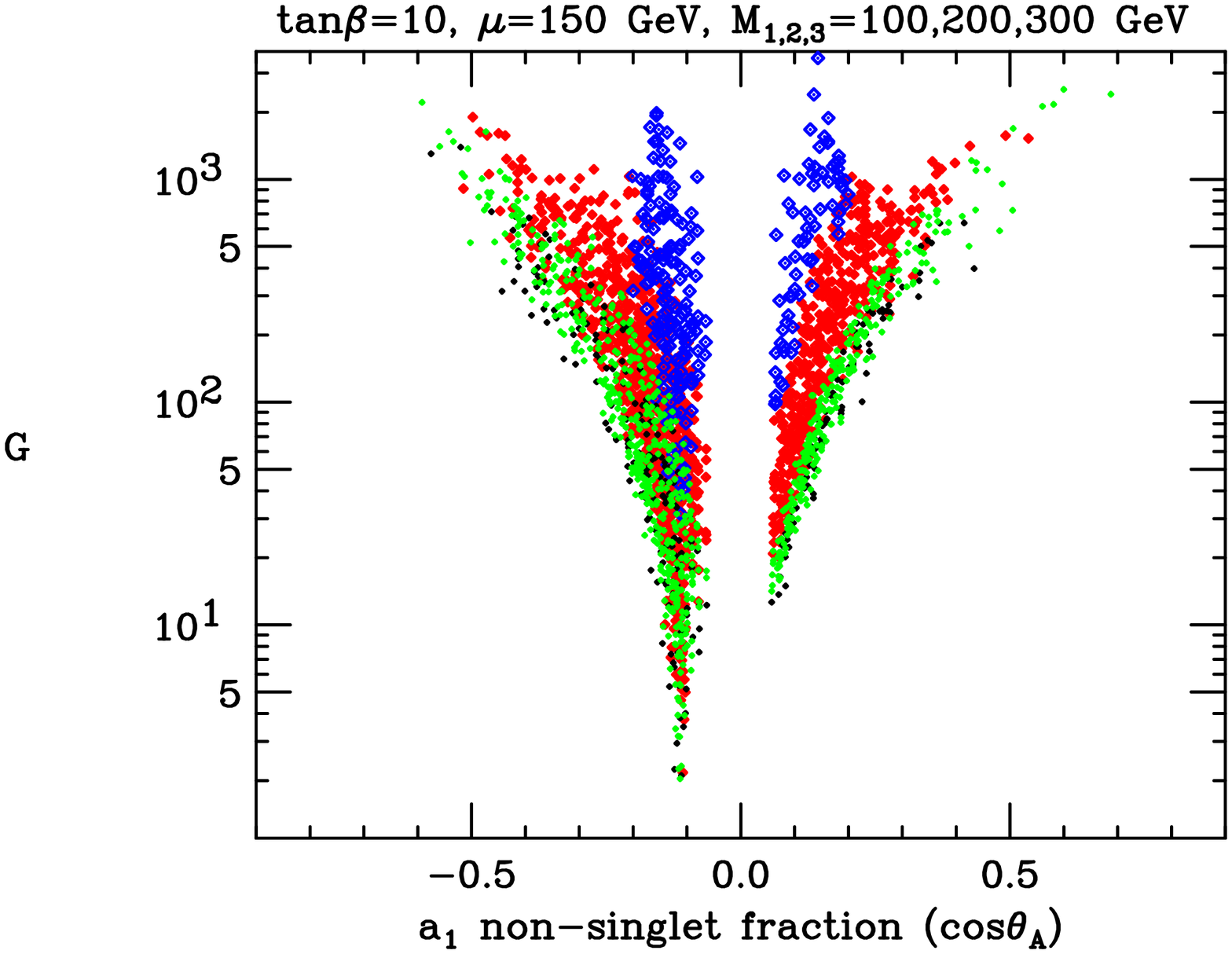}\hspace*{-.5cm}\includegraphics[width=2.5in,angle=90]{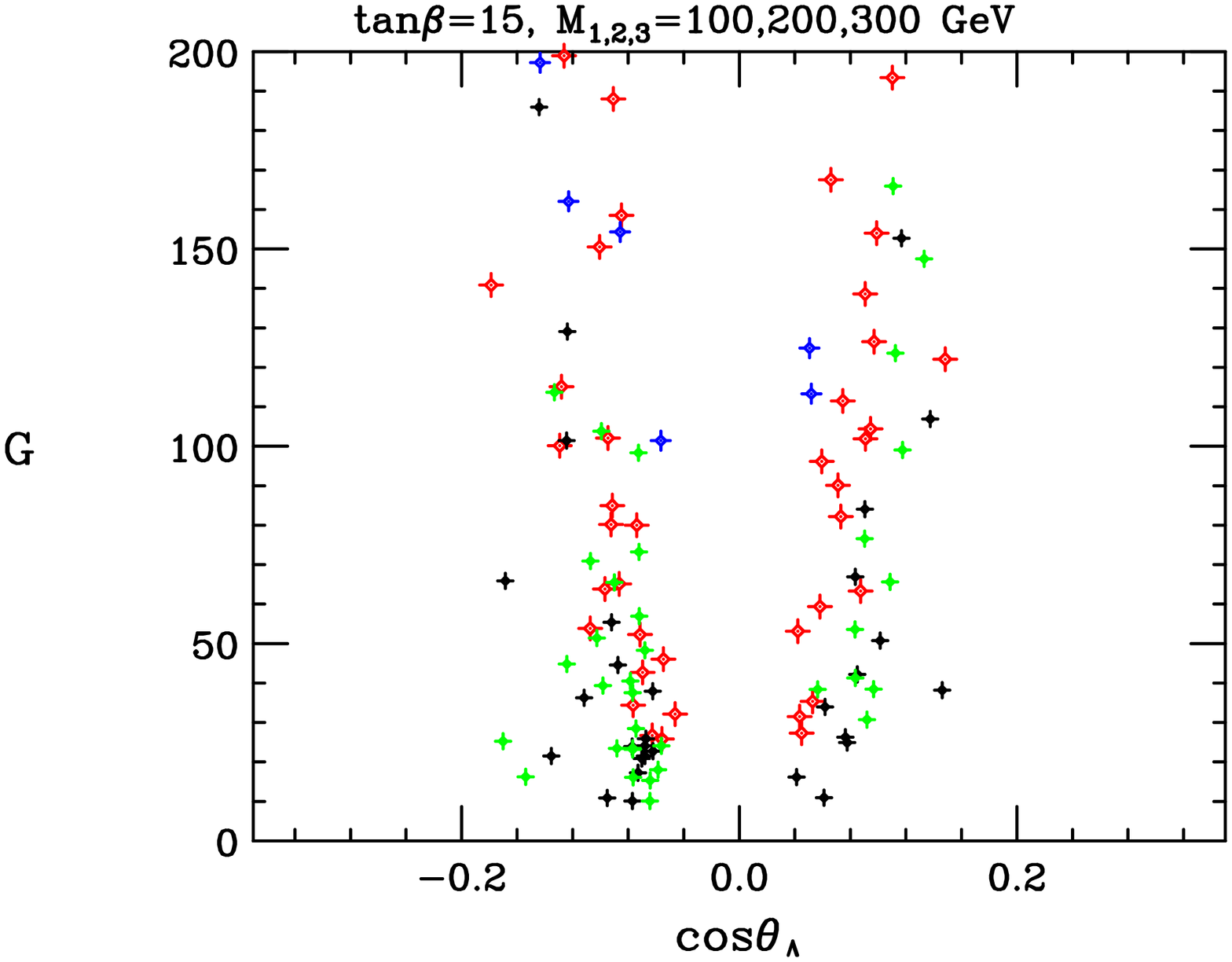}
\caption{ $G$ vs. $\cta$
  for $M_{1,2,3}=100,200,300\gev$ and $\tanb=10$ from $\mueff=150\gev$
  scan (left) and for points with $F<15$ (right) having $\mai<2\mb$
  and large enough $\br(\hi\to\ai\ai)$ to escape LEP limits. The color
  coding is: blue = $\mai<2\mtau$; red $=2\mtau<\mai<7.5\gev$; green
  $=7.5\gev<\mai<8.8\gev$; and black $= 8.8\gev<\mai<9.2\gev$.  }
\label{gvscta}
\end{figure}

Quantitatively, we measure the tuning needed to get small $\mai$ and
large $\br(\hi\to\ai\ai)$ using a quantity called $G$ (the
"light-$\ai$ tuning measure"). We want small $G$ as well as small $F$
for scenarios such that the light Higgs is consistent with LEP limits. Fig.~\ref{gvsf}
shows that it is possible to get small $G$ and small $F$
simultaneously for phenomenologically acceptable points if
$\mai>2\mtau$ (but still below $2m_b$).
A phenomenologically important quantity is $\cta$, the coefficient of
the MSSM-like doublet Higgs component, $A_{MSSM}$, of the $\ai$
defined by
\beq
\ai=\cta A_{MSSM}+\sta A_S\,
\eeq
where $A_S$ is the singlet pseudoscalar field.  The value of $G$ as a
function of $\cta$ for various $\mai$ bins is shown in
Fig.~\ref{gvscta} for points consistent with LEP bounds.
Really small $G$ occurs for $\mai>7.5\gev$ and $\cta\sim -0.1$. Also
note that there is a lower bound on $|\cta|$.  This lower
bound arises because $\br(\hi\to\ai\ai)$ falls below $0.75$ for too
small $|\cta|$.  For the preferred $\cta\sim -0.1$ values, the $\ai$
is mainly singlet and its coupling to $b\anti b$, being proportional
to $\cta\tanb$, is not enhanced.  However, it is also not that
suppressed, which has important implications for $B$ factories.

\section{Detection of the NMSSM light Higgs bosons}

We now turn to how one can detect the $\hi$ and/or the $\ai$.  At the
{\bf LHC}, all standard LHC channels for Higgs detection fail: \eg\ 
$\br(\hi\to\gam\gam)$ is much too small because of large
$\br(\hi\to\ai\ai)$.  The possible new LHC channels are as follows.
\underline{$WW\to \hi\to\ai\ai\to 4\tau$.} This channel looks
moderately promising but complete studies are not available.
\underline{ $t\anti t \hi\to t \anti t\ai\ai\to t\anti t
  4\tau$.} A study is needed.  \underline{$\cntwo\to
  \hi\cnone$ with $\hi\to \ai\ai\to4\tau$.}  This might work given
that the $\cntwo\to \hi \cnone$ channel provides a signal in the MSSM
when $\hi\to b\anti b$ decays are dominant. A $4\tau$ final state
might have smaller backgrounds.  Last, but definitely not least,
\underline{diffractive production $pp\to pp\hi\to pp X$} looks quite
promising.  The mass $\mx$ can be reconstructed with roughly a
$1-2\gev$ resolution, potentially revealing a Higgs peak, independent
of the decay of the Higgs.  The event is quiet so that the tracks from
the $\tau$'s appear in a relatively clean environment, allowing track
counting and associated cuts.  Our \cite{Forshaw:2007ra} results are
that one expects about $3-5$ clean, \ie\ reconstructed and tagged
events with no background, per $30\fbi$ of integrated luminosity.
Thus, high integrated luminosity will be needed.

The rather singlet nature of the $\ai$ and its low mass, imply that
direct production/detection will be challenging at the LHC.  But,
further thought is definitely warranted.

At the {\bf ILC}, $\hi$ detection would be much more straightforward.
The process $\epem\to ZX$ will reveal the
$\mx\sim\mhi\sim100\gev$ peak no matter how the $\hi$ decays.
But the ILC is decades away.

\begin{figure}
\includegraphics[width=3in,angle=90]{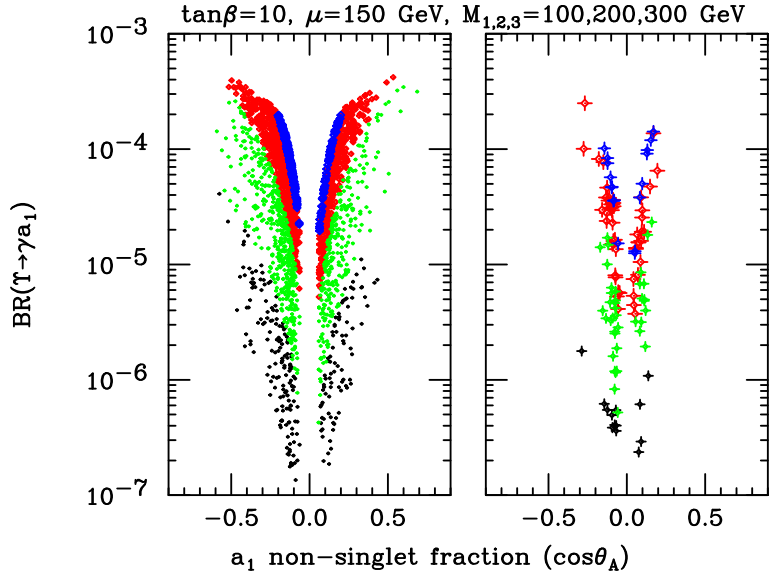}
\caption{$\br(\Upsilon\to\gam\ai)$ for NMSSM scenarios. Results are
  plotted for various ranges of $\mai$ using the color scheme of
  Fig.~\ref{gvscta} (blue, red, green, black correspond to increasing
  $\mai$ in that order).  The left plot comes from an $\alam,\akap$
  scan, holding $\mueff(\mz)=150\gev$ fixed.  The right plot shows
  results for $F<15$ scenarios with $\mai<9.2\gev$ found in a general
  scan over all NMSSM parameters.  The lower bound on $\brups$ arises
  basically from the LEP requirement of $\br(\hi\to\ai\ai)>0.7$ which
  leads to the lower bound on $|\cta|$ noted in text.}
\label{upsilon}
\end{figure}

At {\bf B factories} it may be possible to detect the $\ai$
via $\ups\to\gam \ai$ decays \cite{Dermisek:2006py}.  Both BaBar and
CLEO have been working on dedicated searches.  CLEO has placed some
useful, but not (yet) terribly constraining, new limits. The predicted
values of $\br(\ups\to\gam\ai)$ for $F<15$ NMSSM scenarios are shown
in Fig.~\ref{upsilon}. Note that the scenarios with no light-$\ai$
fine-tuning are those with $|\cta|$ close to the lower bound and
$\mai$ near to $\mups$, implying the smallest values of
$\br(\ups\to\gam\ai)$.
Of course, as $\mai\to \mups$ phase space for the decay causes
increasingly severe suppression.  And, there is the small region of
$\mups<\mai<2\mb$ that cannot be covered by $\ups$ decays.  However,
Fig.~\ref{upsilon} suggests that if $\brups$ sensitivity can be pushed
down to the $10^{-7}$ level, one might discover the $\ai$.
The exact level of sensitivity needed for full coverage of points with $\mai<9.2\gev$
is $\tanb$-dependent, decreasing to a few times $10^{-8}$ for
$\tanb=3$ and increasing to near $10^{-6}$ for $\tanb=50$.
Discovery of the $\ai$ at a $B$ factory would be very important input to the LHC program.

\section{Cautionary Remarks}

The scenario with dominant $\hi\to \ai\ai\to 4\tau$ and $\mhi\sim
100\gev$ certainly has many attractive properties.  However, one can
get quite different scenarios by decreasing the attractiveness
somewhat.  First, one could \underline{relax light-$\ai$ fine-tuning,
  $G$}.  While $\mai<2\mtau$ points have larger $G$ values than points
with $\mai>2\mtau$, we should be prepared for the former possibility.
It yields a very difficult scenario for a hadron collider,
$\hi\to\ai\ai \to 4j$.  Of course, a significant fraction will be
charmed jets.  A question is whether the $pp\to pp \hi$ production
mode might provide a sufficiently different signal from background in
the $\hi\to 4j$ modes that progress could be made.  If the $\ai$ is
really light, then $\hi\to 4\mu$ could be the relevant mode.  This
would seem to be a highly detectable mode, so don't forget to look for
it --- it should be a cinch compared to $4\tau$.  Second, we can
\underline{allow more electroweak / $\mz$-fine-tuning} corresponding
to higher $F$.  In Fig.~\ref{fvsmhinmssm}, the blue squares show that
$\mhi\sim 115\gev$ with $\mai$ either below $2\mb$ or above $2\mb$ can
be achieved if one accepts $F>10$ rather than demanding the very
lowest $F\sim 5$ fine-tuning measure.  Of course, we do not then
explain the $2.3\sigma$ LEP excess, but this is hardly mandatory.
And, $\mhi \sim 115\gev$ is still ok for precision electroweak.  Thus,
one should work on $\hi$ detection assuming: (a) $\mhi\geq 115\gev$
with $\hi\to \ai\ai\to 4\tau$; and (b) $\mhi\geq 115\gev$ with $\hi\to
\ai\ai\to 4b$.  The $pp\to pp \hi$ analysis in case (a) will be very
similar to that summarized earlier for $\mhi\sim 100\gev$, but
production rates will be smaller.  In case (b), there are several
papers in the literature claiming that such a Higgs signal can be seen
~\cite{Carena:2007jk,Cheung:2007sva} in $W\hi$ production.

The most basic thing to keep in mind is that
for a primary Higgs with mass $\lsim 150\gev$,
dominance of $\hi\to\ai\ai$ decays, or even
$\hii\to \hi\hi$ decays, is a very generic feature of any model with
extra Higgs fields, supersymmetric or otherwise.
And, these Higgs could decay in many ways in the most general case. 

Further alternatives arise if there is more than one singlet
superfield.  String models with SM-like matter content that have been
constructed to date have many singlet superfields.  One should
anticipate the possibility of several, even many different Higgs-pair
states being of significance in the decay of the SM-like Higgs of the
model.  Note that this motivates in a very general way the importance
of looking for the light CP-even or CP-odd Higgs states in $\ups\to
\gam X$ decays. 

Another natural possibility is that the $\hi$ could decay to final
states containing a pair of supersymmetric particles (one of which
must be a state other than the LSP if $\mhi<114\gev$).  A particular
case that arises in supersymmetric models, especially those with extra singlets, is $\hi\to \cntwo
\cnone$ with $\cntwo \to f \anti f \cnone$ --- see
\cite{Chang:2007de,Chang:2008cw}.  Once again, the very small $b\anti
b$ width of a Higgs with SM-like couplings to SM particles means that
this mode could easily dominate if allowed.  As noted in
Table~\ref{lepmodes}, LEP constraints allow $\mhi<100\gev$ if this is
an important decay channel.  Higgs discovery would be really
challenging if $\hi\to \ai\ai\to 4\tau$ and $\hi\to \cntwo\cnone\to
f\anti f\ \emiss$ were both present.

\section{Conclusions}

The NMSSM can have small fine-tuning of all types. First, quadratic
divergence fine-tuning is erased ab initio.  Second, electroweak
fine-tuning to get the observed value of $\mz^2$ can be avoided for
$\mhi\sim 100\gev$, large $\br(\hi\to\ai\ai)$ and $\mai<2m_b$.
Light-$\ai$ fine-tuning to achieve $\mai<2\mb$ and (simultaneously)
large $\br(\hi\to\ai\ai)$ (as needed above) can be avoided ---
$\mai>2\mtau$ with $\ai$ being mainly singlet is somewhat preferred to
minimize light-$\ai$ fine-tuning.  Thus, requiring low fine-tuning of
all kinds in the NMSSM leads us to expect an $\hi$ with $\mhi\sim
100\gev$ and SM-like couplings to SM particles but with primary decays
$\hi\to\ai\ai\to 4\tau$.

The consequences are significant.  Higgs detection will be quite
challenging at a hadron collider.  Higgs detection at the ILC is easy
using the missing mass $\epem \to Z X$ method of looking for a peak in
$M_X$.  Higgs detection in $\gam\gam\to \hi\to\ai\ai$ will be easy.
The $\ai$ might be detected using dedicated $\ups\to\gam\ai$ searches.
The stops and other squarks should be light. Also, the gluino and,
assuming conventional mass orderings, the wino and bino should all
have modest mass.  As a result, although SUSY will be easily seen at
the LHC, Higgs detection at the LHC will be a real challenge.  Still,
it now appears possible with high luminosity using doubly-diffractive
$pp\to pp \hi\to pp 4\tau$ events.  Even if the LHC sees the
$\hi\to\ai\ai$ signal directly, only the ILC and possibly $B$-factory
results for $\ups\to\gam\ai$ can provide the detailed measurements
needed to verify the model.

It is likely that other models in which the MSSM $\mu$ parameter is
generated using additional scalar fields can achieve small fine-tuning
in a manner similar to the NMSSM.  However, it is always the case that
low electroweak fine-tuning will require low SUSY masses which in turn
typically imply $\mhi\sim 100\gev$.  Then, to escape LEP limits large
$\br(\hi\to\ai\ai+f\anti f\ \emiss+\ldots)$, with most final states not
decaying to $b$'s (\eg\ $\mai<2\mb$) would be needed.  In general, the
$\ai$ might not need to be so singlet as in the NMSSM and would then
have larger $\brups$.

If the LHC Higgs signal is really marginal in the end, and even if
not, the ability to check perturbativity of $WW\to WW$ at the LHC
might prove to be very crucial to make sure that there really is a
light Higgs accompanying light SUSY and that it carries most of the SM
coupling strength to $WW$.

It is also worth noting that a light $\ai$ allows for a light $\cnone$
to be responsible for dark matter of correct relic density
\cite{Gunion:2005rw}: annihilation would typically be via
$\cnone\cnone\to\ai$.  To check the details, properties of the $\ai$
and $\cnone$ would need to be known fairly precisely.  The ILC might
be able to measure their properties in sufficient detail to verify
that it all fits together.  Also $\ups\to\gam\ai$ decay
information would help tremendously.

In general, as reviewed in \cite{Chang:2008cw}, the Higgs sector is
extraordinarily sensitive to new physics from some extended model
through operators of the form $H^\dagger\ H E$, where $H$ is a SM or
MSSM Higgs field and $E$ is a gauge singlet combination of fields from
the extended sector such a $\phi^\dagger\phi$ or $\phi+\phi^\dagger$,
where $\phi$ is a singlet scalar field from the new physics sector. In
the former case, the operator will have a dimensionless coupling
coefficient and in the latter case a dimensionful coupling
coefficient.  This implies that in either case this new operator is
likely to have large impact on Higgs decays.  In the NMSSM, the
supersymmetric structure implies a slightly more complex arrangement:
the superpotential component $\lam\what S \what H_u\what H_d$ and
soft-SUSY-breaking term $\lam\alam SH_uH_d$ both establish a
connection between the MSSM sector and the extended singlet field
sector and lead to large modifications of the light Higgs decays.
Ref.~\cite{Chang:2008cw} reviews other proposals for the extended
sector. In some, $E$ has higher dimensionality and the operator
coupling coefficient is suppressed by the new physics scale but
nonetheless would greatly influence Higgs physics.  In general, the
precision electroweak preference for a Higgs $h$ with SM-like $WW,ZZ$
couplings and $\mh\sim 100\gev$ greatly increases the odds that a
SM-like Higgs is present but decays to new physics channels. In this
context, SUSY is strongly motivated since electroweak fine-tuning is
minimized precisely for $\mh\sim 100\gev$ and an extended SUSY model
such as the NMSSM can provide the needed non-SM Higgs decays.
\vspace*{-.2in}

\begin{theacknowledgments}
This work is supported in part by the U.S. Department of Energy.  I am
grateful to the Kavli Institute for Theoretical Physics for support
during the project.  Most importantly, I would like to thank George
Rupp for the opportunity to present this overview and honor Mike
Scadron on his 70th birthday in the process.

\end{theacknowledgments}

\vspace*{-.2in}
\bibliographystyle{aipproc}   % if natbib is available

\bibliography{scadron}

\end{document}